# Terahertz Channel performance in ULEO Satellite-to-Ground Communications

**Mingxia Zhang,[1] Wanzhu Chang,[1,2] Yang Jie,[1] Hong Liang,[3,*] Yiming Zhao,[2,4] Houjun Sun,[1,2] Jianjun Ma[1,2,*]**

*[1]School of Integrated Circuits and Electronics, Beijing Institute of Technology, Beijing, 100081, China*
*[2]Tangshan Research Institute, Beijing Institute of Technology, Tangshan, Hebei, 063099 China*
*[3]Radar Operation Control Department, China Meteorological Administration, Beijing, 100044, China*
*[4]School of Interdisciplinary Science, Beijing Institute of Technology, Beijing, 100081 China*

**Abstract:** The exponential growth in satellite data traffic demands communication systems exceeding current microwave capacity limitations, while the terahertz (THz) frequency band (0.1-10 THz) offers unprecedented bandwidth potential with superior weather resilience compared to optical systems, particularly when combined with ultra-low Earth orbit (ULEO) satellite deployments below 300 km altitude. This article presents a comprehensive performance evaluation for ULEO-THz satellite-to-ground communications, analyzing three distinct transmission architectures —direct satellite-to-ground (S2G), satellite-relay-ground (SRG) forwarding, and satellite-to-high altitude base station (S2H) with fiber backhaul. Our analysis leverages altitude-resolved atmospheric propagation models validated using year-long meteorological data from four high-altitude stations in Tibet and Qinghai, China. It incorporates frequency-dependent atmospheric absorption using ITU-R standards, free-space path loss with curved atmospheric modeling, and regional atmospheric variations to derive total channel path loss, available bandwidth capacity, and bit error rate (BER) performance under both AWGN and Weibull fading conditions across multiple THz frequencies. Results demonstrate that direct S2G transmission at lower THz frequencies achieves optimal practical performance with maximum available bandwidth under QPSK modulation, while SRG suffers prohibitive cumulative losses from multiple hops, and S2H is rendered impractical for long-haul links by substantial electro-optical conversion and fiber transmission losses.

## 1. Introduction

Satellite-to-Ground (S2G) communications constitute the fundamental infrastructure enabling global connectivity across remote terrestrial regions, maritime environments, and airborne platforms, serving as critical enablers for applications ranging from disaster response coordination to military operations [1, 2]. Microwave frequency allocations, including Ka and Ku bands, limit data rates to below 100 Gbps [3, 4], bottlenecking high-demand applications, like HD-EO systems, telemedicine platforms, and inter-satellite networks [5, 6]. Exponential satellite deployment growth has further caused spectrum congestion, increasing co-channel interference and degrading link quality [7], thereby undermining RF communication system efficiency and reliability. Laser-based optical communication systems demonstrate high-throughput capabilities exceeding 1.73 Tbps (terabit-per-second) through exploitation of multi-terabit optical bandwidths [3, 8]. However, optical channels exhibit critical vulnerabilities to atmospheric disturbances [9], including tropospheric turbulence causing intensity scintillation effects and stratospheric aerosol distributions contributing to signal fading [10]. Solar background radiation could introduce shot noise that degrades receiver sensitivity and overall system performance [11], while extremely stringent pointing accuracy requirements below 0.001° [12] impose severe mechanical stability constraints on satellite payload design and operation [13, 14].

The terahertz (THz) frequency band (0.1-10 THz) represents an underexploited resource with superior potential for satellite communications. Theoretical analyses establish that THz systems deliver at least one order of magnitude higher data rates compared to millimeter-wave architectures [6]. Beyond raw bandwidth, the THz band offers intrinsic physical advantages, including high directivity enabling secure links, fine spatial resolution facilitating dense spatial multiplexing [15]. Through advanced signal processing techniques including OFDM and spatial multiplexing schemes, this band enables Tbps transmission [16, 17]. Critically, THz channels demonstrate superior propagation characteristics in adverse weather conditions, including fog, haze, and moderate cloud cover, compared to optical channels [18], stemming from their significantly narrower water vapor absorption line profiles in the THz spectrum [19] and substantially reduced susceptibility to atmospheric turbulence-induced distortions [20]. Experimental validation confirms THz systems achieve markedly higher channel availability than optical systems under moderate rainfall and foggy conditions [21, 22]. Despite these advantages, THz implementation faces several technical challenges including limited output power from even quantum cascade lasers and Gunn diodes, constraining EIRP levels [23], and high noise figures in even superconducting mixer receivers requiring elevated signal-to-noise ratio [24]. Additionally, severe atmospheric attenuation from water vapor absorption limits terrestrial transmission distances, though these limitations are substantially mitigated in upper atmospheric and near-space environments where low atmospheric density creates favorable THz channel propagation characteristics.

The deployment of ultra-low Earth orbit (ULEO) satellites, operating at altitudes below 300 km, establishes optimal conditions for THz-based satellite-to-ground (S2G) communication systems through fundamental improvements in link geometry and atmospheric channel propagation characteristics. ULEO configurations achieve substantial reductions in S2G propagation distances, typically decreasing transmission paths by 40%-60% relative to conventional Low Earth Orbit (LEO) architectures [25], together with reduced free space path loss and minimized communication latency [26]. These characteristics enable significant reductions in satellite development and launch costs while simultaneously improving information transmission efficiency, establishing critical advantages for space science exploration and environmental monitoring applications. Contemporary satellite deployment programs have demonstrated the operational viability of ULEO systems. SpaceX's Starlink v2.0 constellation maintains stable operations within the 335-345 km orbital range through advanced ion propulsion systems, while Amazon's Kuiper project targets deployment of 3236 satellites at 350 km altitude for global coverage [27]. The European Space Agency's Very Low Earth Orbit (VLEO) initiative operates Earth observation satellites at 200-250 km altitudes, maintaining orbital stability through sophisticated aerodynamic stabilization technologies [28]. The convergence of ULEO orbital mechanics with THz communication capabilities creates unprecedented opportunities for ultra-high-throughput S2G communication systems, leveraging both the reduced transmission paths and the inherent bandwidth advantages of THz frequencies.

Comprehensive characterization of S2G channel performance represents a critical prerequisite for successful deployment of THz communication systems in satellite applications. Pioneering investigations established THz satellite-to-aircraft communication as a viable research domain, with systematic analysis of propagation characteristics demonstrating theoretical feasibility of space-to-ground THz channels achieving data rates between 50-150 Gbps [6]. Comparative analysis between conventional Ku-band (18 GHz) systems and THz (220 GHz) architectures revealed superior capacity performance, with 220 GHz systems delivering 22-31 Gbps higher throughput over 1000 km transmission distances; implementation of multi-relay network architectures achieved 90% reduction in effective link distances, substantially enhancing overall S2G system performance [29]. Advanced atmospheric channel modeling studies utilizing the LBLRTM computational framework demonstrated transmittance exceeding 98.3% at 99 km altitude, enabling 9.25 THz usable bandwidth over 2 km propagation distances - representing a dramatic improvement over the 173 GHz bandwidth achievable at sea level over 100 m distances, thereby confirming the fundamental superiority of high-altitude environments for THz S2G applications [30]. Atmospheric channel modeling in Ali, Tibet revealed strong seasonal variability in THz channel propagation [31], even though it was limited by its reliance on vertical path assumptions and the absence of end-to-end performance evaluation. Further investigation of propagation characteristics at discrete frequencies (140, 300, 750, and 875 GHz) demonstrated that high-altitude platform-to-satellite links operating between 19 km and 100 km altitudes achieved complete utilization of the 1-1000 GHz frequency spectrum with 80 dBi antenna gain configurations, resulting in 7.7% path loss reduction compared to conventional Ku-band systems [32].

To harness the potential of ULEO-THz links, several transmission architectures have been proposed, each addressing specific propagation and deployment challenges. The direct Satellite-to-Ground (S2G) link serves as the baseline architecture, offering a single hop with minimal infrastructure complexity but traversing the entire atmosphere [6, 33, 34]. The Satellite-Relay-Ground (SRG) architecture, employing an intermediate aerial relay, can potentially mitigate link blockage caused by obstacles or severe weather in the lower troposphere and extend coverage areas [35, 36]. Alternatively, the Satellite-to-High altitude platform (S2H) architecture, which uses a high-altitude ground station (often situated in dry, high-altitude regions) with fiber-optic backhaul, aims to minimize tropospheric attenuation by situating the receiving terminal above a significant portion of the absorbing atmosphere [31, 37]. This study selects these three architectures - S2G, SRG, and S2H - for comparison as they represent fundamentally different approaches to managing the dominant challenges in ULEO-THz links: atmospheric absorption (S2H), path loss and coverage (SRG), and system complexity (S2G).

However, existing studies have not yet systematically compared ULEO-THz transmission channels with different architectures to identify optimal configurations, lack regional atmospheric characterization using real-world long-term meteorological data for channel performance estimation, and have not established the practical frequency selection criteria and modulation schemes necessary for achieving Tbps capacities in operational ULEO-THz S2G systems. This article addresses the knowledge gaps in ULEO-THz communication systems through the analysis of THz channel propagation in ULEO scenarios, quantitative comparison of three distinct ground access topologies (S2G, SRG, and SH), and altitude-resolved atmospheric modeling validated using meteorological data from a full year of China's high-altitude stations. It derives path loss for all these S2G architectures, evaluates achievable bandwidth capacity for ULEO-THz links, assesses bit error rate (BER) performance under available operational conditions, and concludes with comprehensive implications for future system deployment strategies.

The following of this article is organized as follows: Section II details the system model, encompassing three distinct transmission configurations and the channel propagation model. Section III derives the channel performance for the S2G channel utilizing this system model. Subsequently, Section IV evaluates the available bandwidth of the ULEO-THz channel, followed by Section V, which assesses its BER performance. Concluding remarks are provided in the final section.

## 2. Channel models

The channel modeling framework encompasses three distinct communication architectures for ULEO satellite systems, as depicted in Fig. 1. These configurations represent fundamental approaches to addressing the unique propagation challenges inherent in THz satellite communications: direct S2G transmission providing immediate connectivity between satellite platforms and terrestrial terminals; SRG forwarding architecture incorporating satellite-to-relay transmission with subsequent relay-to-ground distribution via established

LTE infrastructure; and S2H configuration utilizing fiber backhaul connectivity, wherein high-altitude platforms function as intermediate nodes with fiber-optic backhaul links to low-elevation terminal networks [38]. Channel performance degradation in these ULEO-THz systems results from three primary attenuation mechanisms: molecular absorption losses within the atmospheric transmission medium, fundamental free-space path loss (FSPL), and inherent communication system losses [39]. Given that all three architectures incorporate satellite-to-low-altitude transmission segments, the foundational S2G channel model serves as the analytical basis for subsequent system comparisons. The modeling framework adopts an atmospheric curvature principle, with ground-based stations positioned at sea level reference altitude, high-altitude platforms deployed at 5 km elevation, and relay nodes strategically positioned at 11 km altitude to optimize coverage and minimize atmospheric attenuation effects.

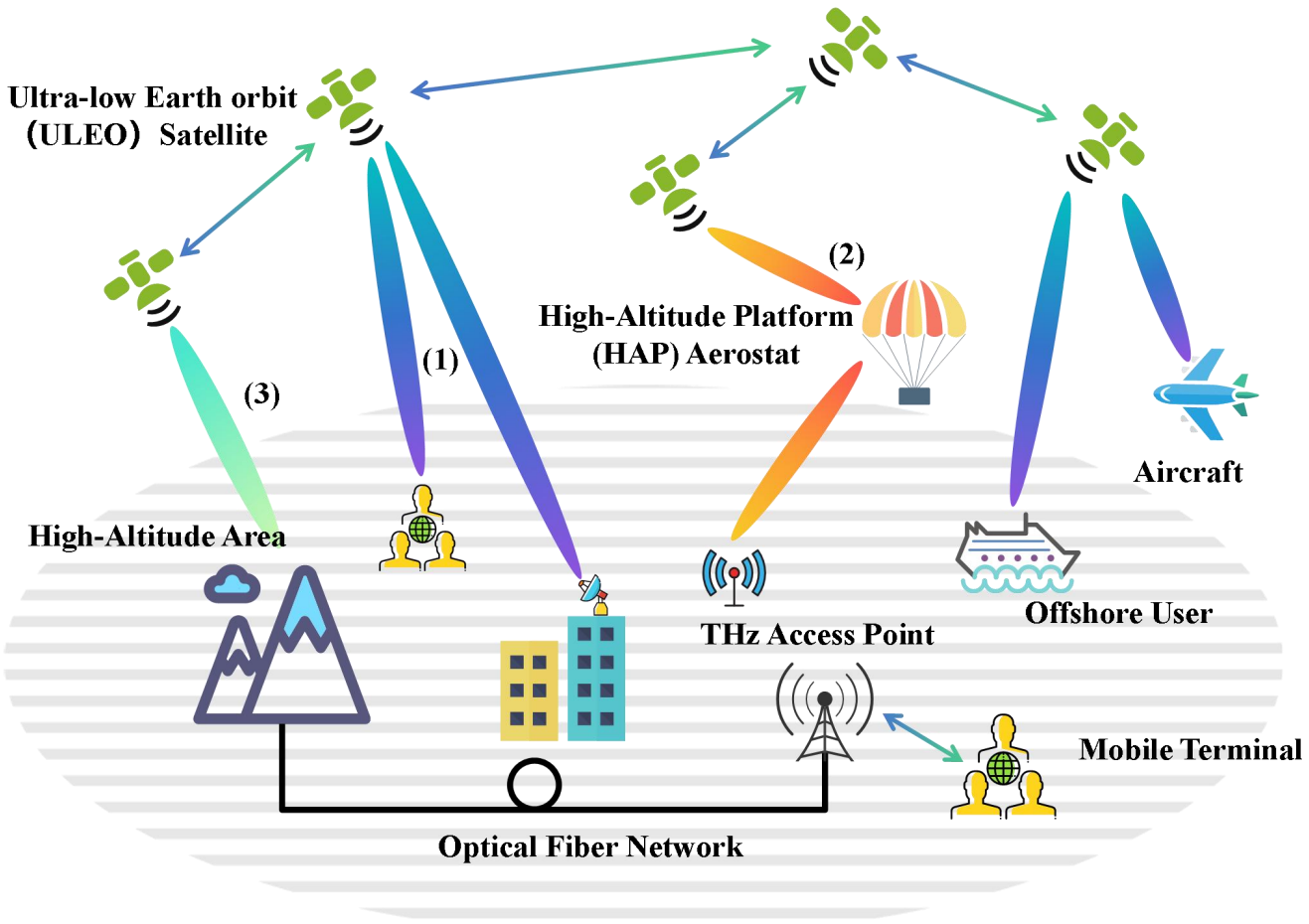

**Fig. 1.** Three ULEO-THz communication architectures: (a) direct Satellite-to-Ground (S2G), (b) Satellite-Relay-Ground (SRG) forwarding, and (c) Satellite-to-High altitude base station (S2H) with fiber backhaul.

The free-space path loss represents the fundamental channel resulting from spherical wavefront spreading in free space [40]. Based on established Friis transmission theory, as $L_{FS}(r,f)=(4\pi r/\lambda)^2=(4\pi fr/c)^2$, with $c$ being the speed of light in vacuum, $\lambda$ the operational wavelength, and $r$ the slant range between the satellite transmitter and receiver. Such inherent frequency-squared dependence characteristic of THz channels introduces substantial attenuation penalties compared to conventional microwave systems. Over 1 km transmission distance, a 140 GHz channel suffers 19.4 dB higher path loss relative to a 15 GHz Ku-band implementation, significantly exceeding conservative satellite link margin specifications [41, 42] and necessitating advanced antenna architectures for viable system operation. To address these fundamental propagation challenges, the proposed S2G communication system employs dual Cassegrain reflector antenna configurations with 0.82 m effective aperture diameter, achieving 60 dBi directional gain [43, 44] at 140 GHz operational frequency through optimized 0.70 illumination efficiency design parameters. This high-gain antenna approach represents a critical system requirement for overcoming the substantial free-space path loss penalties inherent in THz S2G communication links.

Atmospheric absorption constitutes the dominant propagation impairment in THz S2G channels, primarily governed by molecular absorption phenomena from atmospheric oxygen ($O_2$) and water vapor ($H_2O$) constituents. The ITU-R P.676-13 recommendation [19, 45] provides the authoritative framework for quantifying these frequency-dependent attenuation mechanisms through specific attenuation coefficients. The total frequency-dependent specific attenuation coefficient $\gamma(f)$, expressed in dB/km, incorporates dual atmospheric absorption components, as

$$\gamma(f)=\gamma_o(f\ \rho\ T)+\gamma_w(f,\rho\ T)=0.1820f\left(N''_{O_2}(f\ \rho\ T)+N''_{wv}(f\ \rho\ T)\right), \tag{1}$$

where $\gamma_o$ denotes the absorption coefficient due to dry air, which exhibits dependence on atmospheric pressure $p$ (in hPa) and temperature $T$ (in K), and the absorption coefficient $\gamma_w$ due to water vapor, influenced by water vapor density $\rho$ (in g/m$^3$) and ambient temperature $T$, with $f$ representing the carrier frequency in GHz.

The computation of total atmospheric path loss $L_{atm}$ requires integration along the complete propagation trajectory through the atmospheric medium, as

$$L_{atm}(h,f)=\int_{h_a}^{h_s}\left[\gamma_o\left(f,\rho(h),T(h)\right)+\gamma_o\left(f,\rho(h),T(h)\right)\right]dh \tag{2}$$

accounting for altitude-dependent variations in atmospheric density and composition. This integration process considers the total altitude differential along the atmospheric transmission path, where $h$ represents the altitude parameter extending from ground level to satellite orbital position. The mathematical framework incorporates the non-uniform atmospheric conditions encountered throughout the vertical channel propagation path. Additional atmospheric impairments introduce significant performance degradation under adverse meteorological conditions. Precipitation effects manifest as substantial attenuation penalties, with rainfall rates of 8 mm/hr generating specific attenuation coefficients reaching 8 dB/km at 300 GHz [46]. Water cloud contributes supplementary attenuation ranging from 0.1-0.5 dB/km at 100 GHz [47], while atmospheric turbulence phenomena induce scintillation effects that further degrade signal quality [20]. Quantitative analysis demonstrates that at 410 GHz operational frequency under conditions of 20$^{o}$C temperature and 40% relative humidity, strong turbulence conditions reduce bandwidth efficiency to 64.76% over 1 km propagation distances [48], illustrating the critical importance of atmospheric modeling accuracy for practical THz S2G system deployment.

The characterization of atmospheric absorption suffered by THz S2G channels necessitates detailed analysis of the altitude-dependent atmospheric parameters, encompassing temperature gradients, pressure variations, and molecular density profiles throughout the vertical atmospheric column. Fig. 2 presents the vertical distribution of atmospheric absorption loss for a 140 GHz channel extending from sea level to the upper atmospheric boundary at 100 km altitude, computed according to the ITU-R P.835-6 reference standard atmospheric model [49] (detailed atmospheric profiles are provided in Supplementary Information), alongside corresponding relative humidity distributions. The selection of 140 GHz operational frequency reflects optimal considerations for long-range THz link implementation, leveraging the substantial output power capability of 250 mW [50] and achieving high receiver sensitivity of -69.7 dBm under standardized conditions of 1 GHz system bandwidth and 10 dB demodulation signal-to-noise ratio [50, 51]. The curves demonstrate that absorption loss coefficient exhibits exponential decay characteristics with increasing altitude. The absorption loss coefficient decreases from 1.04 dB/km at Earth's surface to 0.004 dB/km at 11 km altitude, becoming negligible. This distribution pattern reflects fundamental atmospheric physics, including barometric air density reduction, diminishing tropospheric water vapor concentrations, and stratospheric temperature inversion phenomena, which collectively render the upper atmosphere essentially transparent to THz channel. It also confirms that atmospheric absorption contributions from 0-11 km altitude dominate total path loss, while contributions above 11 km remain negligible for practical channel performance evaluations.

The atmospheric structure analysis (based on ITU-R P.835-6 standard atmospheric model [52]) in Fig. 2(b) reveals the tropopause boundary at 11 km altitude, above which the stratospheric environment exhibits extremely low density and water vapor content, with corresponding sharp reductions in relative humidity. Under standard atmospheric conditions, cumulative attenuation reaches 1.77 dB for complete vertical transmission paths. The substantial attenuation reduction achieved through elevated transmission initiation points becomes evident when comparing cumulative losses: transmission commencing at 5 km altitude accumulates only 0.1210 dB total loss to 100 km (see Fig. 2(a)), representing a dramatic improvement over sea-level initialization.

Validation through regional atmospheric analysis utilizes measured meteorological data from the Yangbajing station to quantify local atmospheric variations and their impact on THz channel propagation. Fig. 2(c) presents calculated atmospheric absorption loss distribution based on measured and averaged relative humidity data from the Yangbajing region throughout July 2024, as shown in Fig. 2(d), with meteorological measurements provided by the China Meteorological Administration and processed through statistical interpolation techniques. The analysis of peak annual humidity conditions demonstrates atmospheric absorption decreasing from 0.17 dB/km at 5 km altitude to approximately $10^{-6}$ dB/km at 15 km elevation, yielding 0.2477 dB total integrated loss. This represents a 0.1267 dB excess over standard ITU-R reference atmospheric conditions, directly attributable to elevated local relative humidity during the summer monsoon period. These findings underscore the critical importance of implementing region-specific and season-specific atmospheric modeling approaches [19] rather than relying exclusively on global standard atmospheric profiles. This observed deviation, while appearing modest in isolation, becomes operationally significant when extrapolated across long-distance terrestrial transmission paths and wideband THz channel implementations, directly influencing link margin calculations and introducing measurable channel dispersion effects [53].

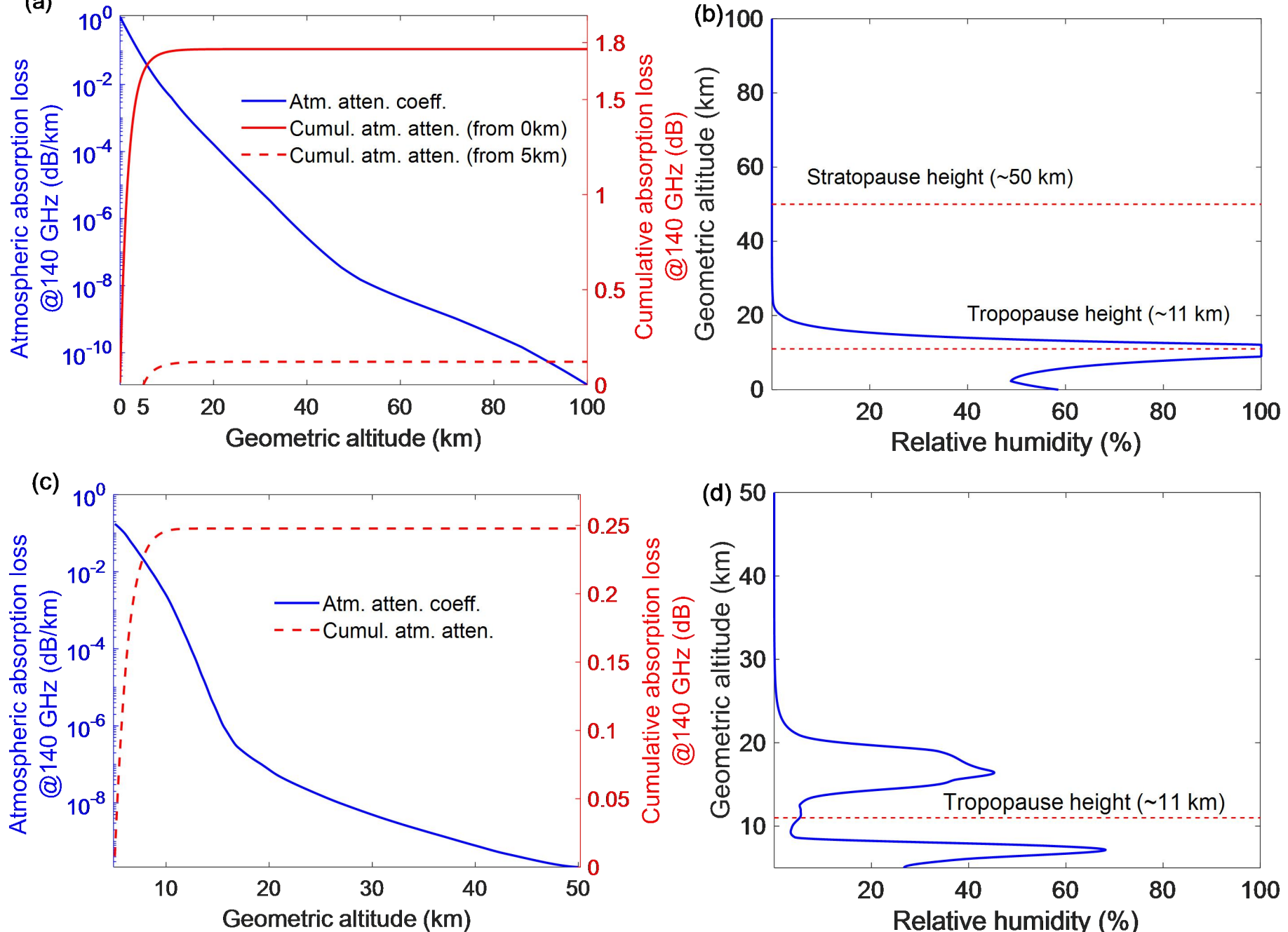

**Fig. 2.** Altitude profiles of atmospheric absorption loss (dB/km) and relative humidity (%) suffered by the 140 GHz channels: (a) absorption absorption loss under standard atmosphere; (b) relative humidity under standard atmosphere; (c) averaged absorption loss from Yangbajing meteorological data throughout July 2024; (d) relative humidity averaged from Yangbajing measurements throughout July 2024.

Reliable THz channel modeling requires recognition that the channel propagation occurs along inclined transmission paths rather than vertical altitude differences, necessitating sophisticated geometric considerations in atmospheric absorption calculations [54]. In this work, we conduct computation of effective path length through the atmospheric medium via integration techniques that incorporate elevation angle effects through the established cosecant law. This is particularly critical for low-elevation-angle transmissions, where extended propagation paths through denser lower atmospheric layers significantly amplify absorption losses. The implementation of a curved atmospheric model [6], addresses the limitations of planar approximations that overestimate atmospheric thickness. The slant range $r_{as}$ between ULEO satellites and ground stations, as

$$r_{as} = \sqrt{\left(R + h_a\right)^2 + \left(R + h_s\right)^2 - 2\left(R + h_a\right)\left(R + h_s\right)\cos(\delta)} \tag{3}$$

incorporates Earth's curvature through geometric relationships involving Earth's radius $R$, ground station altitude $h_a$, satellite altitude $h_s$, and the angular separation parameter ($\delta$) between the ground station and satellite looked from the center of the Earth. The critical atmospheric propagation distance ($r_{atm}$) extends from the ground station through the atmospheric boundary, as

$$r_{atm} = \left(R + h_a\right)\cos\left(\varepsilon\right) + \frac{1}{2}\sqrt{\left(-2\left(R + h_a\right)\cos(\epsilon)\right)^2 - 4\left(\left(R + h_a\right)^2 - b^2\right)}, \tag{4}$$

calculated using the sum ($b$) of Earth's radius and total atmospheric thickness. The parameter $\epsilon$ is the angle between line from Earth to ground station and the line from ground station to satellite. Following molecular absorption effects integration, the resultant atmospheric loss ($L_{atm}$) incorporates these geometric corrections for enhanced accuracy, as

$$L_{atm} = \int_{r_1}^{r_2} \left[ \gamma_{\mathrm{o}}\left(f, \rho(h), T(h)\right) + \gamma_{\mathrm{w}}\left(f, \rho(h), T(h)\right) \right] dr_{atm} \ . \tag{5}$$

Computational precision enhancement utilizes atmospheric layer discretization with 10-meter resolution intervals, providing superior accuracy compared to previous methodological approaches [6, 31]. Earth's curvature effects progressively reduce THz channel propagation distances with increasing altitude within each atmospheric layer. Atmospheric parameters representing each layer's midpoint altitude serve as characteristic values for power loss calculations, with total power loss derived through cumulative summation across all discrete layers. The comprehensive total loss formulation combines free-space path loss with atmospheric absorption loss through superposition principles, as

$$L_{total}\left(r, f\right) = L_{FS}\left(r, f\right) + L_{atm}\left(r, f\right), \tag{6}$$

with the parameters $L_{FS}$ and $L_{atm}$ defined above.

Theoretical validation employs hourly meteorological data from four high-altitude Chinese regions spanning the complete 2024 calendar year, provided by the China Meteorological Administration. The dataset comprises vertical atmospheric profiles including temperature, humidity, pressure, and water vapor density. The original data cover 24 vertical levels from local elevation up to 50 km altitude, with 100 interpolated data points generated using cubic Hermite interpolation for atmospheric layer calculations. The analysis encompasses Ali in Tibet (80.026°E, 32.326°N; elevation: 4500 m) [55, 56], Tanggula Station in Tibet (91.9°E, 32.9°N; elevation: 5000 m) [55] , Yangbajing in Tibet (90.55°E, 30.10°N; elevation: 4300 m) [57], and Snow Mountain Pasture in Qinghai (97.38°E, 37.37°N; elevation: 4300 m). As shown in Fig. 3, the atmospheric absorption loss at 30° elevation angle was calculated for these four regions at 140 GHz carrier frequency, covering from their respective local altitudes up to 50 km altitude, with the reason that the atmospheric absorption is negligible at altitude above 15 km in Fig. 2(c). The regional analysis demonstrates remarkable stability throughout annual cycles, with fluctuations constrained within 1 dB margins, which is different with the calculation result as in reference [31]. This stability provides critical advantages for all-weather and all season ULEO-THz communication systems by reducing excessive link budget margin requirements and minimizing seasonal outage risks.

Snow Mountain Pasture in Qinghai exhibits optimal atmospheric conditions with minimal absorption loss variability, establishing it as the premier candidate for ground terminal deployment. This assessment receives independent validation through the Purple Mountain Observatory (Chinese Academy of Sciences)'s 2024 installation of a 15-meter submillimeter-wave telescope at this location, confirming superior atmospheric conditions for both astronomical observations and THz communications applications.

Conversely, the Yangbajing region demonstrates relatively pronounced monthly variations, particularly during summer periods driven by humidity fluctuations. This makes it an appropriate representative case study, as its meteorological data capture the worst-case seasonal variability that THz links must be designed to withstand.

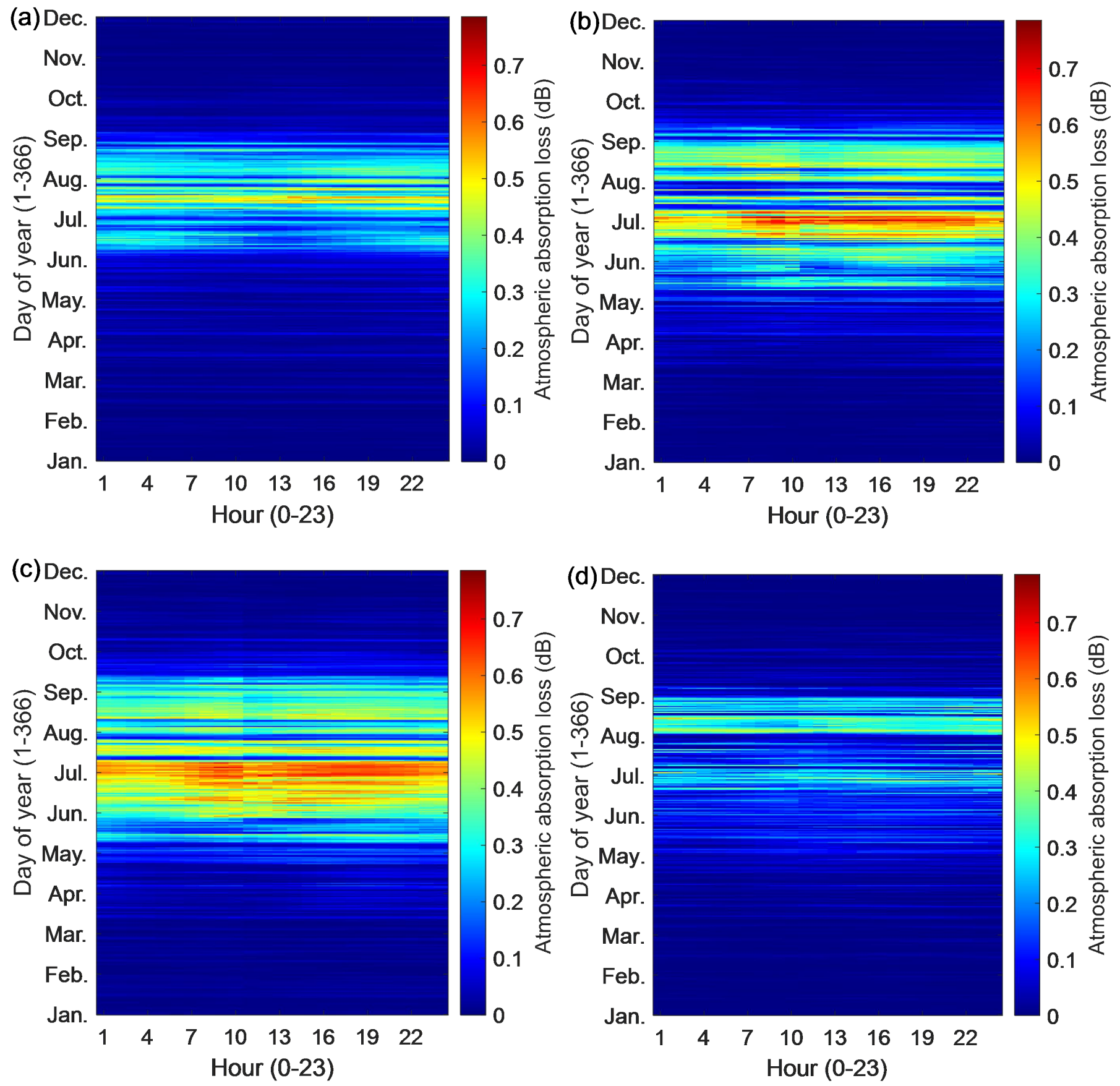

**Fig. 3.** Annual atmospheric absorption loss suffered by the 140 GHz channel for 30° elevation S2G links from four high-altitude stations: (a) Ali, (b) Tanggula, (c) Yangbajing, and (d) Snow Mountain Pasture.

## 3. Channel performance

The evaluation of THz channel performance encompasses all the three distinct transmission architectures (S2G, SRG and S2H) for ULEO S2G satellite systems operating at 30° elevation angles. The selection of 30° elevation angle reflects empirical operational data demonstrating that Starlink satellite constellation elevation angle probability density peaks within the 25°-30° range [58], establishing this parameter as representative of realistic deployment scenarios. We incorporate the unique topographical and atmospheric characteristics of the Yangbajing ground station in Tibet, providing region-specific atmospheric absorption and channel path loss calculations under representative high-altitude conditions.

Figure 4 compares the three architectures. Under idealized propagation, the S2H link achieves the lowest channel path loss, as it initiates transmission from an altitude of 5 km, thereby bypassing the most absorptive lower troposphere (Fig. 4b). However, this channel advantage is fundamentally negated by prohibitive system-level penalties. The model incorporates a 25-30 dB electro-optical conversion loss [59, 60] and a standard single-mode fiber (SMF-28) attenuation of 0.158 dB/km [61]. For a long-haul distance of 1000 km, the fiber loss alone contributes ~158 dB. While optical amplification (e.g., Erbium-Doped Fiber Amplifiers with gains of 20-35 dB [62]) is feasible, compensating for the total attenuation exceeding 183 dB would necessitate a complex, multi-stage amplification chain. This would introduce significant cumulative noise, cost, and power consumption, rendering the S2H architecture economically and practically unviable for long-haul deployment compared to the direct link.

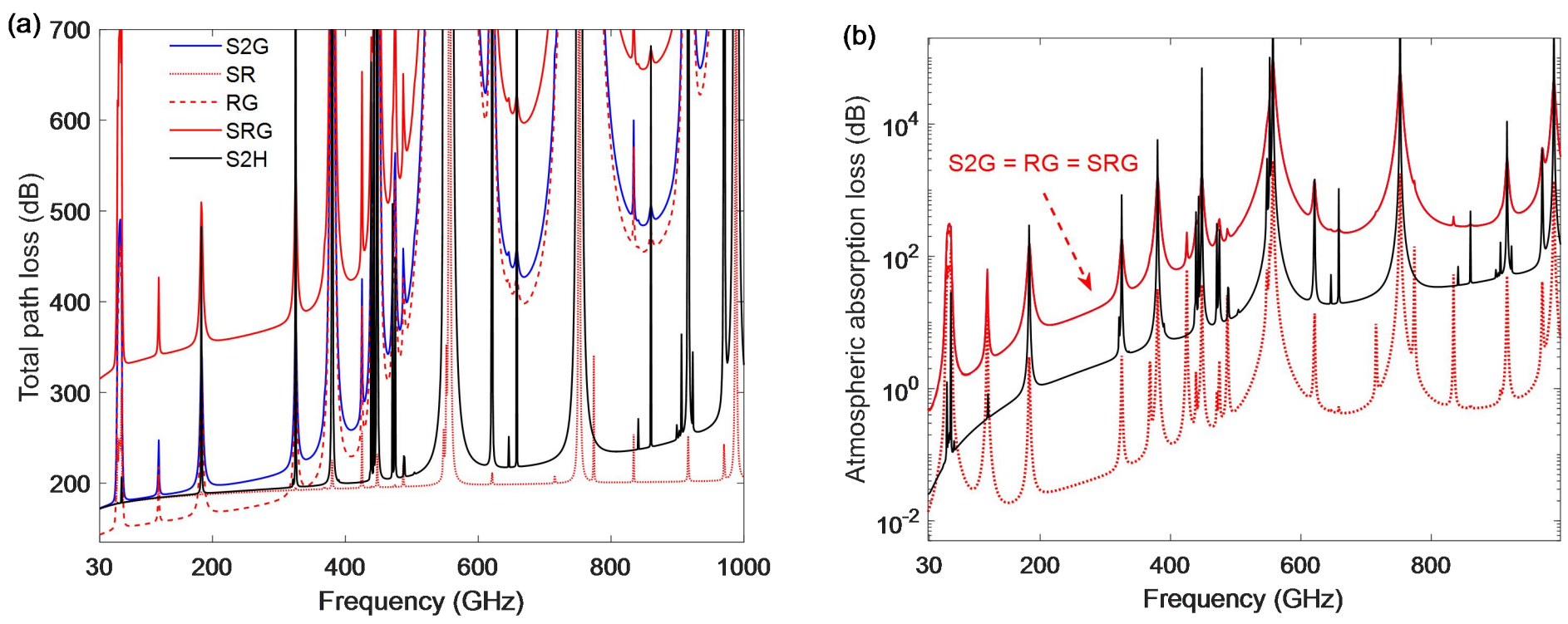

**Fig. 4.** Comparison of (a) total path loss and (b) atmospheric absorption loss for three transmission strategies. SRG = SR+RG and (b) keeps the same legend with (a).

The SRG scheme exhibits the highest total path loss, exceeding that of S2G by approximately 146 dB at 140 GHz. Contrary to initial intuition, this severe degradation is not primarily due to atmospheric absorption, which is nearly identical across S2G, SR, and RG segments (Fig. 4b). Instead, the dominant penalty stems from the fundamental scaling of free-space path loss (FSPL) across multiple hops. Splitting a single ~300 km hop into Satellite-to-Relay (SR, ~289 km) and Relay-to-Ground (RG, ~11 km) segments dramatically increases the sum of FSPL components. While regenerative relays [63] can prevent analog noise accumulation, they do not alleviate the fundamental EIRP requirement for each independent hop. The aggregate attenuation shown here reflects the immense total transmit-power burden, which is prohibitive for power-constrained satellites and relay platforms. This underscores the clear superiority of direct S2G links for long-range THz systems.

Frequency-dependent analysis at 140, 220, and 340 GHz (Table 1) further clarifies design tradeoffs. RG segments (0-11 km) exhibit strong frequency dependence, around 3.53 dB at 140 GHz, rising to 33.33 dB at 340 GH, due to water vapor absorption, marking 140 GHz as the most practical near-ground band. In contrast, SR segments (11-300 km) show negligible absorption (<0.1 dB) across all bands, with FSPL dominating (185-193 dB). Combined SRG losses thus reach 335-390 dB, confirming the RG segment as the primary bottleneck.

For S2G, total attenuation is 189 dB at 140 GHz, 199 dB at 220 GHz, and 226 dB at 340 GHz. A crossover frequency near 315 GHz marks the transition from FSPL-dominated to absorption-dominated loss regimes. The S2H case, modeled for Yangbajing conditions (5-300 km), reduces near-tropospheric contributions, yielding 185-197 dB total loss - better than S2G but still constrained by practical fiber penalties.

**Table 1. Link budget analysis for S2G communication at 140, 220 and 340 GHz.**

| Schemes | Freq. (GHz) | Atmospheric attenuation (dB) | FSPL (dB) | Total attenuation (dB) |
|---|---|---|---|---|
| | 140 | 3.5271 | 156.220 | 159.749 |
| RG (0km-11 km) | 220 | 9.2739 | 160.146 | 169.422 |
| | 340 | 33.3283 | 163.927 | 197.256 |
| | 140 | 0.0214 | 185.060 | 185.084 |
| SR (11km-300 km) | 220 | 0.0281 | 188.986 | 189.016 |
| | 340 | 0.0754 | 192.767 | 193.845 |
| | 140 | 3.5485 | 341.280 | 334.833 |
| SRG (Sum) | 220 | 9.3020 | 349.132 | 358.438 |
| | 340 | 33.4037 | 356.694 | 390.101 |
| | 140 | 3.5271 | 185.382 | 188.911 |
| S2G (0km-300 km) | 220 | 9.2739 | 189.308 | 198.583 |
| | 340 | 33.3283 | 193.089 | 226.417 |
| | 140 | 0.4950 | 185.252 | 185.747 |
| S2H (5km-300 km) | 220 | 1.2752 | 189.178 | 190.453 |
| | 340 | 3.5281 | 192.959 | 196.487 |

## 4. Available bandwidth

The analytical derivation of total available bandwidth for the three communication strategies requires comprehensive signal-to-noise ratio (SNR) analysis incorporating system-specific parameters and propagation characteristics. The fundamental SNR relationship establishes the performance baseline through noise power calculations adhering to established methodologies [30], as

$$\gamma = P_{Tx} + G_{Tx} + G_{Rx} - L_{total} - P_n \tag{7}$$

utilizing a standardized subcarrier bandwidth of 1 GHz. While the HITRAN database integrated within the ITU-R P.676 model maintains spectral resolution finer than 0.005 $cm^{-1}$ [64], our analysis adopts 1 GHz frequency resolution to balance computational efficiency with analytical precision. This bandwidth-to-frequency resolution equivalence enables investigation of multiple discrete channels, each maintaining precisely 1 GHz bandwidth across the considered THz spectrum, regardless of total available bandwidth extent.

Successful communication requirements mandate that achieved SNR values satisfy minimum threshold $\gamma_{th}$ criteria, as

$$\gamma \geq \gamma_{th} \tag{8}$$

expressed through the relationship between maximum permissible path loss and system parameters. The maximum permissible path loss is derived as

$$L_{total}^{th} = P_{Tx} + G_{Tx} + G_{Rx} - \gamma_{th} - P_n \tag{9}$$

The total available bandwidth calculation incorporates usable bandwidth contributions from individual spectral bands meeting threshold requirements, as

$$W_{total} = \sum_i W_i, \forall i \text{ where } L_{toatl} < L_{total}^{th}, \tag{10}$$

with $W_i$ representing the usable bandwidth in the $i$-th spectral band satisfying $L_{total} < L_{total}^{th}$, with system parameters in channel modeling detailed in Table 2. This analytical framework establishes the foundation for comparative performance evaluation across different transmission architectures and modulation schemes.

**Table 2. System Parameters in Channel Modeling.**

| Parameter | Symbol | Value | Unit |
|---|---|---|---|
| Tx Power | $P_{\mathrm{Tx}}$ | 40 | dBm |
| Transmit antenna gain | $G_{\mathrm{Tx}}$ | 60 | dBi |
| Receive antenna gain | $G_{\mathrm{Rx}}$ | 60 | dBi |
| Bandwidth | $B$ | 1 | GHz |
| System noise temperature | $T_{\mathrm{sys}}$ | 290 | K |
| Noise power | $P_n$ | -83.9 | dBm |
| Orbital height | $h$ | 300 | km |
| Elevation angle | $\theta$ | 30 | degree |

Two modulation schemes are analyzed for their contrasting characteristics - QPSK, valued for relative robustness under adverse conditions [65], and 16-QAM, offering higher spectral efficiency [66]. For QPSK modulation operating over additive white Gaussian noise (AWGN) channels, bit energy-to-noise power spectral density ratio (Eb/N0) simulations utilizing MATLAB R2024a *bertool* function establish performance baselines. Achieving BER = $10^{-5}$ requires Eb/N0 of 9.58 dB (SNR ≈ 11.98 dB) for QPSK and 13.43 dB (SNR ≈ 19.45 dB) for 16-QAM.

Theoretical results indicate that QPSK tolerates channel attenuation up to 232.6 dB, compared with 225.0 dB for 16-QAM. Integration with the path loss results in Fig. 4(a) enables calculation of sub-threshold bandwidths across architectures, summarized in Table 3. These results confirm the practical dominance of QPSK. Its lower SNR requirement aligns with the noise-limited nature of satellite channels, while higher-order schemes demand margins rarely achievable in THz S2G environments [67].

The frequency-dependent performance, by employing the results in Fig. 4(a), further guides band selection. At 140 GHz, QPSK achieves the best tradeoff between available bandwidth and attenuation. The 220 GHz band remains viable but requires tighter link adaptation under humid or low-elevation conditions. At 340 GHz, QPSK is marginal and higher-order schemes are infeasible, confining this band to specialized high-altitude deployments. These findings underscore the need for adaptive strategies, such as dynamic modulation switching and/or frequency agility, to sustain reliable THz S2G channels across varying atmospheric conditions.

**Table 3. Available bandwidth comparison for three transmission strategies.**

| Communication schemes | QPSK (GHz) | 16-QAM (GHz) |
|---|---|---|
| S2G | 284 | 262 |
| SR | 913 | 908 |
| S2H | 594 | 557 |

## 5. BER performance

Evaluation of BER performance across different modulation schemes provides critical insights into channel property influences on communication link reliability under realistic operational conditions [68]. Building upon established S2G communication modulation frameworks, this analysis conducts comparative assessment of QPSK performance under varying channel conditions and fading environments. For unobstructed line-of-sight transmission scenarios, system performance analysis employs the classical additive white Gaussian noise (AWGN) model, which provides the theoretical foundation for BER evaluation. The theoretical BER expression under QPSK modulation, as

$$BER_{QPSK} = \frac{1}{2} erfc\left(\sqrt{\frac{E_b}{N_0}}\right) = \frac{1}{2} erfc\left(\sqrt{\frac{\gamma}{2}}\right), \tag{11}$$

establishes the baseline performance relationship through the bit energy-to-noise power spectral density ratio ($E_b/N_0$) of $\gamma$ $(dB)$ = $E_bN_0$ $(dB)$ + 10 lg ($\log_2 M$). The realistic characterization of THz S2G channel propagation necessitates consideration of atmospheric fading phenomena that significantly influence link reliability. The Weibull distribution emerges as a particularly effective model for characterizing multipath fading in environments where traditional statistical distributions prove insufficient [69]. This distribution demonstrates exceptional flexibility in modeling non-homogeneous scattering environments and heavy-tailed amplitude fluctuations [70, 71]. Thus, the Weibull fading model becomes particularly relevant for THz S2G communications traversing atmospheric layers with varying density, composition, and turbulence characteristics.

The probability density function (*pdf*) of instantaneous signal-to-noise ratio in Weibull fading channels incorporates shape and scale parameters ($\beta$, $\lambda$) that control fading severity and distribution characteristics respectively, as

$$f_\gamma(\gamma) = \frac{\beta}{\lambda}\left(\frac{\gamma}{\lambda}\right)^{\beta-1} \cdot \exp\left(-\left(\frac{\gamma}{\lambda}\right)^{\beta}\right), \quad \gamma \geq 0 \,. \tag{12}$$

Here, the shape parameter $\beta$ governs fading intensity, with smaller values indicating more severe fading conditions, while the scale parameter $\lambda$ determines the distribution extent. Average BER calculation requires integration of AWGN BER expressions over the complete fading distribution, as

$$BER_{Weibull} = \int_0^\infty Q(\sqrt{\gamma}) f_\gamma(\gamma) d\gamma \,, \tag{13}$$

typically necessitating numerical methods or approximation techniques due to mathematical complexity. Monte Carlo simulation validation provides empirical verification of theoretical predictions, yielding shape parameter values of $\beta$=0.3398 and scale parameter values of $\eta$=3.4987 through comprehensive data fitting procedures.

The calculation results presented in Fig. 5(a) and 5(b) demonstrate distinct performance characteristics under AWGN and Weibull fading conditions respectively. AWGN channel analysis reveals predictable and stable QPSK performance with strong resilience under line-of-sight dominated scenarios. However, Weibull fading introduction produces significant BER performance degradation, particularly at reduced SNR levels. This degradation aligns with the physical reality of THz channel propagation through atmospheric layers where aerosol scattering, turbulence effects, and localized humidity gradient variations induce stochastic fluctuations in received power [19]. These atmospheric phenomena create challenging propagation environments that substantially differ from idealized free-space conditions.

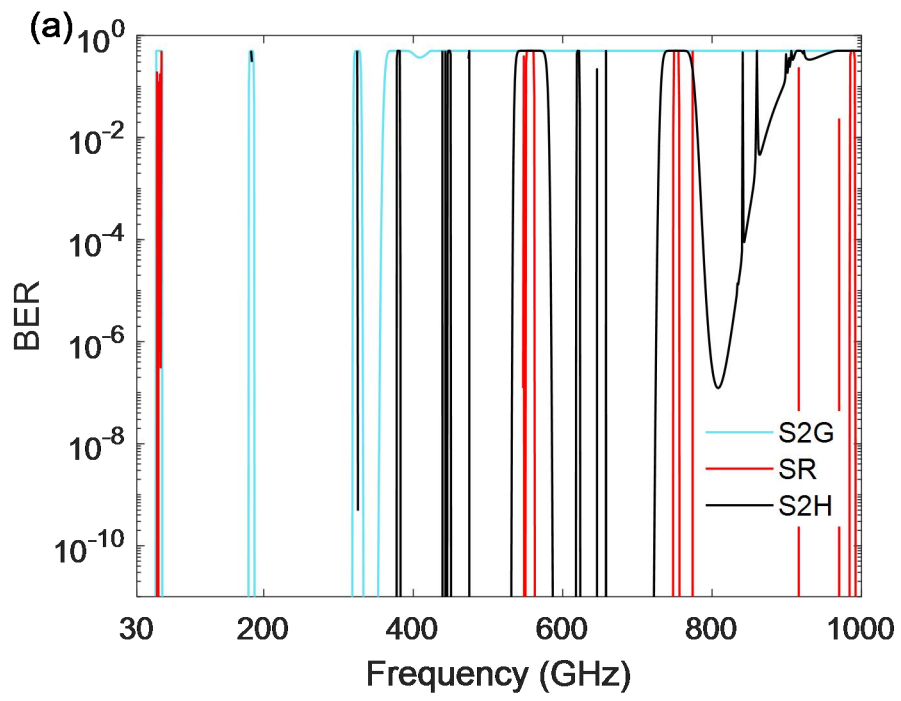

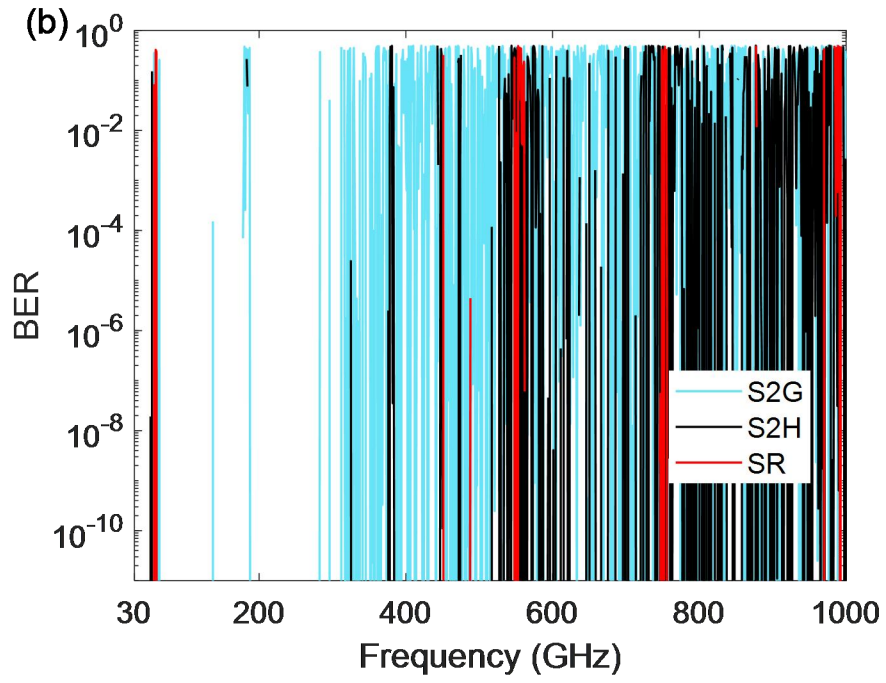

**Fig. 5.** BER performance comparison under QPSK modulation for (a) AWGN and (b) Weibull fading channels.

The analysis across all three communication architectures reveals that the SR configuration achieves optimal BER performance under both AWGN and Weibull fading conditions. This superior performance stems from strategic bypassing of dense tropospheric layers where atmospheric absorption and multipath effects reach maximum intensity. The S2G scheme demonstrates elevated attenuation sensitivity and increased fading vulnerability, consistent with the substantial total attenuation values documented in Table 1, particularly at higher operational frequencies where S2G channels accumulate 226.4 dB attenuation at 340 GHz compared to 196.5 dB for S2H configurations. Both SR and S2H architectures effectively mitigate fading severity through elevated operational altitudes, thereby maintaining broader bandwidth availability even under challenging Weibull fading conditions.

For high-velocity ultra-low Earth orbit satellites (orbital velocity ~7.7 km/s), Doppler shift is a factor that must be considered[60]. At carrier frequencies of 140 GHz and 340 GHz, the maximum Doppler shift is approximately 3.6 MHz and 8.8 MHz, respectively. Although the relative bandwidth occupied by this shift is small compared to the 1 GHz system bandwidth considered in this study, its absolute value far exceeds the acquisition range of conventional phase-locked loops (PLLs). Consequently, the receiver must rely on dedicated digital signal processing (DSP) algorithms for frequency offset estimation and coarse compensation, followed by fine carrier synchronization performed by the PLL to ensure proper system operation. On the other hand, THz communications rely on high-gain antennas (60 dBi in this model), which are highly sensitive to pointing errors caused by satellite attitude jitter, orbital uncertainties, and wind-induced sway of ground stations [72, 73]. Such errors are commonly modeled by a Rayleigh distribution, introducing a multiplicative random fading component $h_m \approx A\exp(-2r^2 / w_{eq}^2)$ [74]. The resulting power loss in dB follows $L_{pointing} \propto (\theta_e / \theta_{3dB})^2$. Although pointing error is a critical consideration in system design, this study focuses on establishing baseline performance bounds for the three architectures under comparison. Assuming high-precision pointing mechanisms (e.g., with error ≤ 0.03°, corresponding to ~0.4 dB loss) are employed, the associated loss is significantly smaller than the path and absorption losses analyzed here (typically >180 dB). Therefore, in this quasi-static link budget analysis, pointing error is not modeled as a random variable, but is instead accounted for within the overall system implementation margin.

## 6. Conclusion

The critical bandwidth limitations of conventional satellite communication systems, combined with severe spectrum congestion from exponential satellite deployment growth, necessitate revolutionary approaches to achieve ultra-high-throughput requirements. This article investigates THz channel performance for ultra-low Earth orbit (ULEO) satellites to demonstrate the fundamental viability of THz frequencies for satellite-to-ground communications through systematic analysis of three transmission architectures. Direct satellite-to-ground (S2G) communication at 140 GHz emerges as the optimal practical configuration, achieving 189 dB total attenuation and 284 GHz available bandwidth under QPSK modulation. While satellite-to-high altitude base station (S2H) configurations achieve superior atmospheric channel characteristics with 186 dB attenuation, electro-optical conversion penalties and fiber transmission losses render this approach impractical for operational deployment. Multi-hop satellite-relay-ground (SRG) forwarding suffers prohibitive cumulative losses, confirming direct transmission superiority despite higher atmospheric absorption. Regional atmospheric modeling using year-long meteorological data from four high-altitude Chinese stations (Ali, Tanggula, Yangbajing, and Snow Mountain

Pasture) revealed remarkable temporal stability with fluctuations within 1 dB, enabling reliable all-weather operations without excessive link margins.

Frequency-dependent analysis across 140, 220, and 340 GHz bands established 140 GHz as optimal, with higher frequencies demonstrating marginal feasibility due to severe water vapor absorption exceeding 33 dB in tropospheric propagation. BER evaluation under AWGN and Weibull fading confirmed QPSK superiority for noise-limited satellite channels. Atmospheric contributions from 0-11 km altitude dominate total attenuation, while stratospheric propagation above the tropopause remains essentially transparent to THz channels, emphasizing ground terminal elevation importance. These findings establish theoretical and practical foundations for terabit-per-second satellite systems, demonstrating that ULEO-THz architectures overcome fundamental constraints of conventional microwave satellite communications through optimal frequency selection, direct transmission paths, and strategic ground station positioning.

**Funding.** National Natural Science Foundation of China under Grant (62471033), and the Special Program Project for Original Basic Interdisciplinary Innovation under the Science and Technology Innovation Plan of Beijing Institute of Technology under Grant (2025CX11010).

**Disclosures.** The authors declare no conflicts of interest.

**Data availability.** Data underlying the results presented in this paper are not publicly available at this time but maybe obtained from the corresponding author - Jianjun Ma upon reasonable request.

**Supplementary Information.** See Supplemental_document for supporting content.